# Enhancing Kinematics Understanding through a Video Game Based on Real-Time Motion Graphs

Mateo Dutra[1], Marcos Abreu[2], Martín Monteiro[3], Silvia Sguilla[4], Cecilia Stari[5], , Álvaro Suárez[6], Arturo Martí[1]

[1]Instituto de Física, Facultad de Ciencias, Universidad de la República, Iguá 4225, CP 11400, Montevideo, Uruguay.

[2]CeRP del Centro, Consejo de Formación en Educación, Administración Nacional de Educación Pública, Independencia y 24 de abril, CP 94000, Florida, Uruguay.

[3]Facultad de Ingeniería, Universidad ORT Uruguay, Cuareim 1451, CP 11100, Montevideo, Uruguay.

[4]Instituto de Formación Docente de Canelones, Consejo de Formación en Educación, Administración Nacional de Educación Pública, Treinta y Tres 470, CP 90000, Canelones, Uruguay.

[5]Instituto de Física, Facultad de Ingeniería, Universidad de la República, Av. Julio Herrera y Reissig 565, CP 11300, Montevideo, Uruguay.

[6]Instituto de Profesores Artigas, Consejo de Formación en Educación, Administración Nacional de Educación Pública, Av. Libertador 2025, CP 11800, Montevideo, Uruguay.

*E-mail: monteiro@ort.edu.uy

Abstract: Interpreting kinematic graphs remains a significant challenge in physics education. The MissionMotion Project addresses this issue by providing a gamified physical-computational environment combining low-cost sensors, physical activity, computational thinking, and real-time visualization of motion graphs. This paper presents the design, development, and implementation of the project, with a particular focus on the pilot phase conducted with high school students in Uruguay. During this phase, we primarily used the MEEGA+ questionnaire to evaluate the gaming experience, usability, and motivation of the participants. Our analysis of the results shows high levels of satisfaction, perceived learning, and engagement, supporting the proposal's viability. Finally, we plan to conduct a large-scale conceptual evaluation to analyze how the proposal impacts understanding of kinematic graphs using standardized assessment tools.

# I. INTRODUCTION

Physics education faces a persistent challenge: students' difficulty in constructing coherent mental models of motion. Numerous studies in the field of Physics Education Research (PER) have documented students' struggles in interpreting and connecting graphs of position, velocity, and acceleration as functions of time (McDermott et al., 1987; Beichner, 1994). Common difficulties include confusion between position, displacement, and distance traveled (Jufriadi et al., 2021); challenges in recognizing the slope of a position–time graph as velocity and that of a velocity–time graph as acceleration (McDermott et al., 1987); as well as "pictorial" interpretations of graphs, in which they are read as literal representations of physical trajectories (Beichner, 1994). In response to this scenario, PER has demonstrated that active learning, collaborative work, and the integration of multiple representations—linguistic, mathematical, graphical, and experimental—can significantly enhance conceptual understanding (Hake, 1998; Freeman et al., 2014), while gamification adds a motivational component that promotes persistence and self-regulation. In addition to sensor-based methodologies for teaching kinematics, recent research explores the integration of computational thinking as a way to strengthen students' reasoning about motion and data analysis (Dutra, Suárez, & Marti, 2025).

In Uruguay, the curriculum introduces kinematics in the early years of secondary education; however, experiences that directly link real physical motion with its graphical representation remain scarce. MissionMotion emerges as a response to this gap, aiming to connect students' real physical movement with the graphical representations they study, within an accessible and engaging environment. This setting promotes active and embodied learning activities to reinforce the phenomenon–representation connection (Thornton & Sokoloff, 1998), integrates computational thinking in physics through data sensing, processing, and visualization (Weintrop et al., 2016), and incorporates gamification strategies to enhance motivation and task engagement (Deterding et al., 2011).

# II. MISSIONMOTION DESCRIPTION

MissionMotion is a cross-platform web application that receives real-time data from sensors and displays it graphically alongside a target graph that the player/student must reproduce or imitate. On the screen, a paper ball appears moving over a reference system represented by a ruler. During the motion, it is possible to visualize either its position or its velocity in real time. The target graphs correspond to position–time or velocity–time relationships, and the ball can be moved using a mouse, trackpad, touchscreen, or ultrasonic sensors connected to micro:bit or Arduino devices. Figure 1 shows a screenshot of the game, where the ball, the ruler, the target graph, and the ongoing construction of the student's graph can be observed. Based on the similarity between the graphs, the player is assigned a score.

The proposal includes different levels of difficulty, individual and collaborative participation modes, and an evaluation system based on the degree of correspondence between the student-generated graph and the reference graph. Interaction can be carried out through body movements, direct manipulation of devices, or robotic control, incorporating immediate visual feedback that fosters self-regulation and progressive improvement through iteration.

The use of ultrasonic sensors on platforms such as micro:bit or Arduino broadens the pedagogical possibilities. It allows the design of activities in which students program the boards, configure sensor connections, and use the playful environment as a resource to develop computational thinking. At more advanced stages, it is even possible to integrate robots programmed to reproduce the target motion, thereby enhancing both technical competencies and conceptual understanding of the represented physical phenomenon.

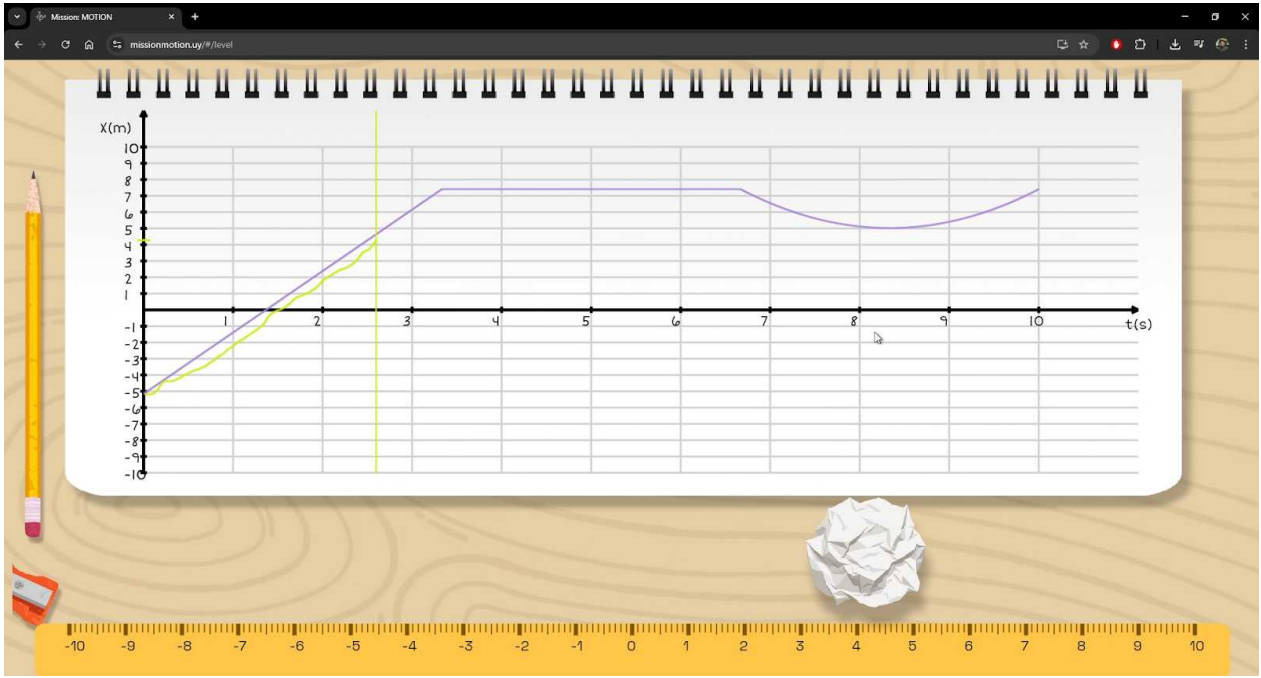

Figure 1: Screenshot of the video game on the difficult level. The purple line is the one that had to be imitated with the movement of the paper ball, which is represented by the green line. The vertical line indicates the current time.

## III. IMPACT RESEARCH METHODOLOGY

In the first stage, we conducted a pilot evaluation with 54 ninth-grade students (14–15 years old) who engaged in activities within the environment during a 2-hour session. The objectives at this stage were to assess the gameplay experience and its motivational potential, identify necessary technical and pedagogical adjustments, and observe interaction dynamics and problem-solving strategies. In these activities, each student worked individually on a computer and controlled the object solely with the mouse. Initially, they played at their own pace, and later participated in a collective session in which they competed at the same levels.

At this stage, we implemented the MEEGA+ questionnaire (Model for the Evaluation of Educational Games) (Petri et al., 2018), an internationally validated instrument that assesses multiple dimensions of usability and gameplay experience in educational environments. Among the aspects evaluated are enjoyment and challenge, perceived learning, ease of use, social interaction (cooperation and competition), as well as immersion and attention. We selected MEEGA+ because it provides a comprehensive diagnosis segmented into specific factors, which guides concrete improvements in both game design and classroom dynamics.

In a second stage, the proposal is planned to be implemented in 4–5 schools with a sample of 150–200 students, under an experimental design with control and experimental groups. In this stage, the focus will not only be on evaluating the tool itself but also on assessing student learning within selected didactic sequences integrated into the environment. For this purpose, standardized tests such as the modified TUG-K (Zavala et al., 2017) will be applied, allowing us to measure students' understanding of kinematics graphs and to evaluate the conceptual impact of the intervention.

## IV. PRELIMINARY RESULTS

In the pilot phase, the results obtained through the MEEGA+ instrument showed high appreciation from the participating students across several key aspects of the proposal. Ninety-two percent of the responses were positive regarding design and aesthetics, highlighting the perception of a visually appealing and motivating environment. The sense of challenge and achievement reached 88% favorable responses, suggesting that the experience was stimulating while remaining attainable for most students. With respect to perceived learning, 85% of the responses indicated that the activity contributed significantly to the understanding of the content. Finally, more than 70% of the students reported that the proposal fostered social interaction, thus reinforcing its potential as a collaborative resource. Figure 2 presents an overall summary of the results.

In the open-ended comments, students emphasized the fun aspect and the clarity of the visual feedback, as well as the motivation generated by the game-like format. They also noted that competitive modes between groups increased participation and engagement. Classroom observations corroborated these perceptions: students were seen deliberately adjusting the movement of the object to improve alignment with the target graph, spontaneous discussions emerged regarding strategies to "beat the score," and, notably, even those students who usually displayed passive participation became actively involved in the activity.

| | Experiencia del Jugador | Totalmente en desacuerdo | No estoy de acuerdo | Indiferente | Estoy de acuerdo | Totalmente de acuerdo | Mediana |
|---|---|---|---|---|---|---|---|
| Confianza | El contenido y la estructura del juego me dan confianza suficiente para reconocer que con el uso del juego puedo aprender algo. | | 4 | 11 | 32 | 8 | 1 |
| Desafío | El juego me plantea retos apropiados. | | 7 | | 33 | 15 | 1 |
| Desafío | El juego ofrece nuevos desafíos (obstáculos, situaciones, variaciones) a un ritmo adecuado. | 5 | 8 | | 33 | 9 | 1 |
| Desafío | El juego no se torna monótono en sus tareas (repetitivo o con tareas aburridas). | 2 | 10 | 21 | 16 | 6 | 0 |
| Satisfacción | Completar las tareas del juego me permitió obtener una sensación de logro. | 1 | 6 | 12 | 16 | 20 | 1 |
| Satisfacción | Gracias a mi esfuerzo personal he avanzado en el juego. | 1 | | 10 | 29 | 15 | 1 |
| Satisfacción | Me siento satisfecho con las cosas que he aprendido con el juego. | | 2 | 19 | 23 | 11 | 1 |
| Satisfacción | Recomendaría el uso del juego a mis amigos. | 4 | 3 | 14 | 23 | 11 | 1 |
| Interacción social | He tenido la oportunidad de interactuar con los demás durante el juego. | 4 | 5 | 12 | 18 | 16 | 1 |
| Interacción social | El juego promueve momentos de cooperación y/o competición entre los jugadores. | 1 | 1 | 13 | 23 | 17 | 1 |
| Interacción social | Me he divertido con otras personas durante el juego. | | 6 | 11 | 23 | 15 | 1 |
| Diversión | Me he divertido. | 1 | 1 | 8 | 28 | 17 | 1 |
| Diversión | Durante el juego hubo algo (elementos del juego, competición, etc.) que me hizo sonreír. | 1 | 5 | 10 | 25 | 14 | 1 |
| Atención centrada | Había algo interesante al comienzo del juego que me llamo la atención. | | 11 | 19 | 18 | 7 | 0 |
| Atención centrada | Mientras estaba concentrado en el juego, el tiempo se ha pasado rápidamente. | 2 | | 20 | 23 | 10 | 1 |
| Atención centrada | Me he sentido totalmente inmerso en la atmósfera del juego, olvidándome incluso de lo que había a mi alrededor. | 6 | 15 | 17 | 10 | 7 | 0 |
| Pertinencia | El contenido del juego es relevante para mis intereses. | 6 | 7 | 21 | 16 | 5 | 0 |
| Pertinencia | El contenido del juego está relacionado con los conocimientos y contenidos de la asignatura. | | | 3 | 24 | 28 | 2 |
| Pertinencia | El uso del juego es un método de aprendizaje apropiado para esta asignatura. | | | 8 | 26 | 21 | 1 |
| Pertinencia | Considero que el uso de este juego es más apropiado para mi aprendizaje que otros métodos dentro de la asignatura. | | 3 | 14 | 23 | 15 | 1 |
| Percepción del Aprendizaje | El juego contribuyó a mi aprendizaje en la asignatura. | 1 | 3 | 16 | 20 | 15 | 1 |
| Percepción del Aprendizaje | En comparación con otras actividades del curso, el juego ha sido eficaz en mi aprendizaje. | 1 | 1 | 20 | 21 | 12 | 1 |
| Percepción del Aprendizaje | El juego me ayudó a comprender las gráficas de posición en función del tiempo. | 2 | 3 | 10 | 18 | 22 | 1 |

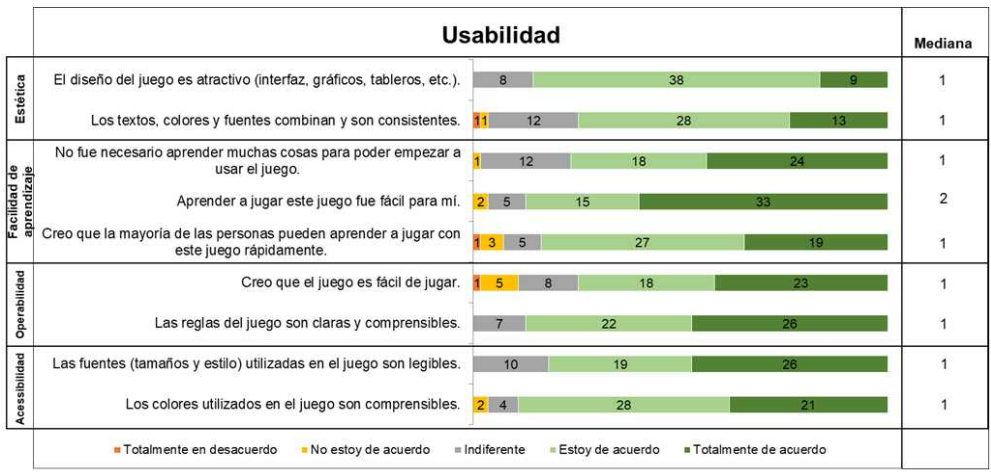

**Figure 2.** Summary of the results from the test application. For each statement, the degree of agreement and the number of students indicating that level are shown using a color code. The median is calculated by scoring responses as follows: −2 for "strongly disagree," −1 for "disagree," 0 for "neutral," +1 for "agree," and +2 for "strongly agree."

## V. DISCUSSION

The application of MEEGA+ in the pilot phase was essential to confirm that *MissionMotion* is not only technically feasible but also fosters high levels of engagement and strong perceptions of learning among students. This instrument made it possible to identify key strengths—such as appealing design, immediate feedback, and the promotion of social interaction—as well as areas for improvement, including difficulty adjustment and sensor calibration. In line with the perspectives of Ainsworth (2006) and Freeman et al. (2014), the results support the idea that the integration of physical activity, multiple representations, and gamification enhances active learning and intrinsic motivation. In the next stage, the incorporation of the modified TUG-K will allow for a quantitative evaluation of the proposal's impact on students' conceptual understanding of kinematics graphs.

## VI. CONCLUSIONS AND FUTURE PROJECTIONS

The *MissionMotion* project stands as an innovative proposal for teaching kinematics, combining sensing, embodied experience, gamification, and immediate feedback within a single learning environment. The preliminary results obtained through MEEGA+ indicate that students not only perceive the experience as attractive and motivating but also value it as a useful tool for improving their understanding of physical concepts. This approach helps break down traditional barriers in teaching, fostering active participation, collaboration, and the integration of multiple forms of representation.

Looking ahead, the potential of *MissionMotion* expands toward broader and more diverse educational contexts. Large-scale implementation across different settings will allow exploration of its adaptability and pedagogical robustness. The inclusion of the modified TUG-K will provide quantitative evidence of its impact on conceptual understanding, while the analysis of the computational thinking involved will open new lines of interdisciplinary research. Finally, the publication of the results and the release of resources in open-access format will ensure that the proposal can be replicated and adapted by the educational community, thus promoting a global movement toward more active, inclusive, and motivating learning experiences.

**Acknowledgments**

This work was supported by the Sectoral Education Fund of ANII and Plan Ceibal. The authors would also like to thank the participating educational institutions and teachers for their collaboration.